\documentclass[british]{article}
\usepackage[T1]{fontenc}
\usepackage[latin9]{inputenc}
\usepackage{geometry}
\usepackage{babel}
\usepackage{amstext}
\usepackage{graphicx}
\usepackage{mathtools}
\usepackage{subfig}
\usepackage[linesnumbered,ruled]{algorithm2e}
\usepackage{esint}
\usepackage[authoryear]{natbib}
\usepackage[unicode=true]{hyperref}
\usepackage{float}
\begin{document}
\title{Rare event ABC-SMC$^{2}$}
\author{Ivis Kerama, Thomas Thorne and Richard G. Everitt}
\maketitle
\begin{abstract}
Approximate Bayesian computation (ABC) is a well-established family of Monte Carlo methods for performing approximate Bayesian inference in the case where an ``implicit'' model is used for the data: when the data model can be simulated, but the likelihood cannot easily be pointwise evaluated. A fundamental property of standard ABC approaches is that the number of Monte Carlo points required to achieve a given accuracy scales exponentially with the dimension of the data. \citet{Prangle2016} proposes a Markov chain Monte Carlo (MCMC) method that uses a rare event sequential Monte Carlo (SMC) approach to estimating the ABC likelihood that avoids this exponential scaling, and thus allows ABC to be used on higher dimensional data. This paper builds on the work of \citet{Prangle2016} by using the rare event SMC approach within an SMC algorithm, instead of within an MCMC algorithm. The new method has a similar structure to SMC$^{2}$ \citep{Chopin2013}, and requires less tuning than the MCMC approach. We demonstrate the new approach, compared to existing ABC-SMC methods, on a toy example and on a duplication-divergence random graph model used for modelling protein interaction networks.
\end{abstract}

\section{Introduction}

\subsection{Approximate Bayesian computation}

This paper concerns Bayesian inference of the parameters $\theta$
of ``implicit'' models: those where the likelihood cannot be evaluated
pointwise at $\theta$. This problem is encountered in a number of
different fields, including epidemiology, ecology, economics, particle physics,
cosmology and genetics. Suppose we wish to use a model $f_{\theta}\left(y\right)$
for observed data $y$ that depends on parameters $\theta$, and wish
to estimate the parameters of the model (sometimes known as \textquotedbl calibrating\textquotedbl{} the model) using Bayesian inference.

Our aim in performing this calibration depends on the problem we are
tackling. Three possible aims are:
\begin{itemize}
\item \textbf{inference} of the parameters $\theta$ or functions of $\theta$,
e.g. the reproduction number $R_{0}$ in an epidemic;
\item \textbf{prediction} of possible future data using the calibrated model,
e.g. number of new infections in an epidemic;
\item \textbf{model criticism or comparison} to help us understand if $f$
is a realistic model, either in terms of prediction, or in terms of
describing the observed data.
\end{itemize}
In each situation the posterior distribution $\pi\left(\theta\mid y\right)$
of parameters $\theta$ plays a central role. For many commonly used
choices of $f$, we can evaluate $f_{\theta}\left(y\right)$ pointwise
at $\theta$. Inference then usually proceeds using a Monte Carlo
method for simulating from the posterior distribution $\pi\left(\theta\mid y\right) \propto p\left(\theta\right)f_{\theta}\left(y\right)$, for example importance sampling (IS), Markov chain Monte Carlo (MCMC) or sequential Monte Carlo (SMC). The standard versions of these algorithms all rely on being able to evaluate the prior and likelihood pointwise at $\theta$, thus cannot be directly implemented for implicit models. This led to the development of approximate Bayesian
computation (ABC), which replaces the likelihood with an assessment as to whether $\theta$ is plausible under the posterior by simulating
$x\sim f_{\theta}\left(\cdot\right)$ and checking if statistics of
$x$ are close to statistics of the observed data $y$. This procedure provides
an approximation to the likelihood (the ``estimated ABC likelihood''),
the use of which results in an approximation to the posterior (``ABC
posterior'').

ABC is a well-established area of research. Much work has focussed
on two issues: how to explore the space of $\theta$ whilst using
the estimated ABC likelihood; and the effect of using the ABC approximation in place of the exact likelihood. Methods for exploring the space of $\theta$ include
the ABC variants of rejection sampling, IS, MCMC, SMC and Bayesian
optimisation: a review can be found in \citet{Fan2019}. The development
of methods for improving the estimation of the likelihood has focussed
on improving both the bias and variance of the estimates. We continue
our introduction by introducing the ABC-MCMC algorithm, and discussing
the bias and variance of the estimated ABC likelihood in this context.

\subsection{ABC-MCMC and the estimated ABC likelihood}

Let $P_{\epsilon}\left(y\mid x\right)$ be the ``ABC kernel'':
a distribution on $y\mid x$, symmetric in $x$, that is specified by normalising a kernel
with tolerance (or bandwidth) $\epsilon\geq0$, which takes larger
values the closer $x$ is to $y$. The ABC-MCMC algorithm of \citet{Marjoram2003}
is shown in algorithm \ref{alg:abc-mcmc}.

\begin{algorithm}
\caption{ABC-MCMC}\label{alg:abc-mcmc}
\KwIn{$N \geq 0$}
\KwOut{$\left\{ \theta^i,x^i\right\}_{i=0}^N$}
 
Initialise $\theta^0$;\

Simulate $x^{*}\sim f_{\theta{^0}} \left(\cdot\right)$;\

\For {$i=1:N$}
{
    Simulate $\theta^{*}\sim q\left(\cdot\mid\theta^{i-1} \right)$;\
    
    Simulate $x^{*}\sim f_{\theta^{*}}\left(\cdot\right)$;\
    
    Let $\left(\theta^i,x^i\right) = \left(\theta^{*},x^{*}\right)$ with probability
    \[
    1\wedge\frac{p\left(\theta^{*}\right)}{p\left(\theta^{i-1}\right)}\frac{q\left(\theta^{i-1} \mid\theta^{*}\right)}{q\left(\theta^{*}\mid\theta^{i-1} \right)}\frac{P_{\epsilon}\left(y\mid x^{*}\right)}{P_{\epsilon}\left(y\mid x^{i-1}\right)};
    \]\
    
    Else let $\left(\theta^i,x^i\right) = \left(\theta^{i-1},x^{i-1}\right)$;\
    
}

\end{algorithm}

One way of understanding ABC-MCMC is as a ``pseudo-marginal'' method
\citep{Beaumont2003,Andrieu2009}. This class of methods uses an unbiased
approximation to the likelihood in place of an exact likelihood in
a Metropolis-Hastings algorithm; it is shown in \citet{Andrieu2009}
that the limiting distribution of the $\theta$-points generated by
such an algorithm is the same as if the exact likelihood were used.
In the case of ABC--MCMC, the likelihood estimate at a point $\theta$
is simply $P_{\epsilon}\left(y\mid x\right)$, where $x\sim f_{\theta}\left(\cdot\right)$.
This is a Monte Carlo estimate of what we call the ``true'' ABC
likelihood
\[
l^{\text{}}\left(y\mid\theta\right)=\int_{x}P_{\epsilon}\left(y\mid x\right)f_{\theta}\left(x\right)dx.
\]
This integral is intractable in general, hence the use of a Monte
Carlo estimator. As pointed out in \citet{DelMoral2012g}, the estimated
ABC likelihood $P_{\epsilon}\left(y\mid x\right)$ is a very high
variance estimate of the true ABC likelihood since it uses only a
single Monte Carlo point, and in some circumstances it may be more
efficient to take the sample average of $P_{\epsilon}\left(y\mid\cdot\right)$
for several simulations from $f_{\theta}\left(\cdot\right)$. In this
case, for $N_{x}$ points simulated from $f_{\theta}\left(\cdot\right)$,
the estimated likelihood is
\begin{equation}
\hat{l}\left(y\mid\theta\right)=\frac{1}{N_{x}}\sum_{n=1}^{N_{x}}P_{\epsilon}\left(y\mid x^{n}\right).\label{eq:abc_llhd_est_multiple}
\end{equation}

For expository purposes, it is useful to view our Monte Carlo estimator
as an IS estimator of the normalising constant $\int_{x}P_{\epsilon}\left(y\mid x\right)f_{\theta}\left(x\right)dx$
of the unnormalised target distribution $P_{\epsilon}\left(y\mid x\right)f_{\theta}\left(x\right)$
when using proposal $f_{\theta}\left(x\right)$. The importance sampling
estimator is unbiased, and its variance of an depends on the distance
between the proposal and the target \citep{Agapiou2015}. \citet{Andrieu2009}
tells us that the unbiasedness of the estimated likelihood will result
in ABC-MCMC having the same invariant distribution as if we had used
the true ABC likelihood.

A likelihood estimator with a higher variance usually results in a
less efficient MCMC algorithm. For the estimated ABC likelihood, we
have that the distance between target $P_{\epsilon}\left(y\mid x\right)f_{\theta}\left(x\right)$
and proposal $f_{\theta}\left(x\right)$ (and hence the variance of
the estimator), will tend to be larger when the dimension of $y$
is higher and when $\epsilon$ is smaller. The dimension of $x$ has
a particularly large impact: the variance of the estimator increases
exponentially with the dimension of $y$. This is the reason that
the raw data is rarely used in ABC: common practice is to reduce the
dimension of the data by using summary statistics, and thus reduce
the variance of the likelihood estimator. If the statistics are not
sufficient, this results in a different posterior to using the full
data: the variance has been reduced at the cost of introducing bias.
A similar tradeoff is made when choosing an appropriate $\epsilon$.
The ABC likelihood will only result in a posterior equal to the true
posterior as $\epsilon \rightarrow 0$, however this is the case where the variance
of the likelihood estimator is at its highest. In practice some $\epsilon>0$
is used, chosen such that the variance of estimates from ABC-MCMC
are not too high, although at the cost of introducing a bias.

The methodology used in this paper is focussed on trying to avoid,
as far as possible, the need to reduce the dimension of the data by
choosing summary statistics, whilst avoiding a high variance likelihood
estimator. The approach we use is described in the following section.

\subsection{Unpacking the black box simulator and using rare event SMC} \label{sec:unpacking}

Many methods for estimating the likelihood can be seen as different
ways of approaching this bias-variance tradeoff: for example some
approaches assume some parametric or non-parametric model for joint
or conditional distributions of $\theta,y$, with the aim of reducing
variance whilst attempting to introduce little additional bias \citep{cranmer_frontier_2020}. Our method
is one in an alternative class of approaches that make use of a decomposition
of the simulator that is available in many situations. We rewrite
the simulator as a deterministic transformation $H$ of the parameter
$\theta$ and a random vector $u$ that is drawn from a tractable
distribution $\phi\left( \cdot \mid \theta \right)$. We believe this ``reparameterisation trick'' was first introduced
in the ABC context in \citet{Andrieu2012f}; it has been used in a number of papers since
(e.g. \citet{Meedsa,Forneron2016,Moreno2016,Graham2017}). The important
property of this decomposition is that it allows us to replace the
target 
\[
P_{\epsilon}\left(y\mid x\right)f_{\theta}\left(x\right)
\]
with
\[
P_{\epsilon}\left(y\mid H\left(u,\theta\right)\right)\phi\left( u \mid \theta \right).
\]
Although we are still using an ABC-style likelihood,
the intractable $f_{\theta}$ is no longer present. This presents
the possibility of exploring other ways of moving around the $\left(\theta,u\right)$-space, for example the Hamiltonian Monte Carlo (HMC) in \citet{Graham2017}). To understand
the possible benefits of using this approach, it is useful to think
again about the reasons for the high variance of the standard ABC
likelihood: in this case the $u$ variable is drawn independently
at each iteration, from a distribution that does not depend on $y$.
We might hope that by tailoring $u$ specific to $\theta$ and $y$
- choosing the random vector $u$ such that our likelihood simulations
conditional on $\theta$ are close to $y$ - we make an efficiency
gain.

In this paper we follow \citet{Prangle2016} in using a "rare event" SMC algorithm
for simulating from the conditional distribution of $u\mid\theta,y$
in order to tailor $u$ to $\theta$ and $y$. In fact, this simulation
of $u$ conditional on $y$ and $\theta$ is used to estimate the
ABC likelihood $l^{\text{}}\left(y\mid\theta\right)=\int_{u}P_{\epsilon}\left(y\mid H\left(u,\theta\right)\right)\phi\left( u \mid \theta \right)du$,
with a lower variance than the standard approach, which we presented
above as importance sampling. This is exactly the marginal particle
MCMC algorithm \citep{Andrieu2010l} in the case of ABC. Theoretical
results about SMC tell us that we expect the variance of this likelihood
estimator to scale more favourably with the dimension of $y$ (which
for simplicity we assume is the same as the dimension of $u$) than
the standard IS based estimator. \citet{Prangle2016} studies theoretically the computation required to accept a proposed point using the IS estimator compared to the SMC as $d\rightarrow \infty$, and finds a cost of $O\left( \epsilon^{-d} \right)$ for the IS approach, compared to $O\left( d^2 \log\left( \epsilon^{-1} \right) \right)$ for SMC (also see \citet{Beskos2014b}).

This approach should not be confused with ABC-SMC algorithms, which
are described later in this paper. These algorithms are focussed on
using SMC to explore $\theta$-space (sometimes based on a justification
of exploring the joint $\left(\theta,u\right)$-space). The SMC method employed by \citet{Prangle2016} is instead exploring the conditional distribution of $u\mid\theta,y$.
When the ABC kernel is a uniform distribution, such an SMC algorithm
has been explored thoroughly in the work of \citet{Cerou2012}, under
the guise of a method for rare event estimation. For the situation of ABC, we use a sequence
of $T$ targets with the final one being $P_{\epsilon}\left(y\mid H\left(u,\theta\right)\right)\phi\left( u \mid \theta \right)$.
The \textquotedbl 0th target\textquotedbl{} (the proposal) is given
by $\phi\left( u \mid \theta \right)$, and the $t$th target (for $1\leq t\leq T$ is $P_{\epsilon_{t}}\left(y\mid H\left(u,\theta\right)\right)\phi\left( u \mid \theta \right)$,
where $\infty>\epsilon_{1}>...>\epsilon_{T}=\epsilon$. The algorithm
then proceeds as in algorithm \ref{alg:rare-event-smc}; our notation is that the values taken by particles
in the SMC sampler have a $\left(\cdot\right){}^{m}$ superscript;
so for example $u_{t}^{m}$ is the $u$-value taken by the
$m$th particle at the $t$th target. $\mathcal{M}\left( \left( w_t^1, ..., w_t^{N_u} \right) \right)$ in the resampling step stands for the multinomial distribution which assigns probability $w_t^n$ to outcome $n \in 1:N_u$. The algorithm mentions that resampling will be performed if some degeneracy condition is met: the most common choice is if the effective sample size, estimated using
\[
\left( \sum_{n=1}^{N_{\theta}} \left(w^n\right)^2 \right)^{-1}
\]
falls below some proportion $\alpha \in (0,1)$ of $N_u$. All SMC algorithms in this paper include a step that normalises the weights. For algorithm \ref{alg:rare-event-smc}, this step uses
\[
w^n_{t} = \frac{\tilde{w}^n_{t}}{\sum_{i=1}^{N_u} \tilde{w}^i_{t}},
\]
for each particle; an analogous form is used in the other algorithms.

\begin{figure}
\centering
\includegraphics[scale=1]{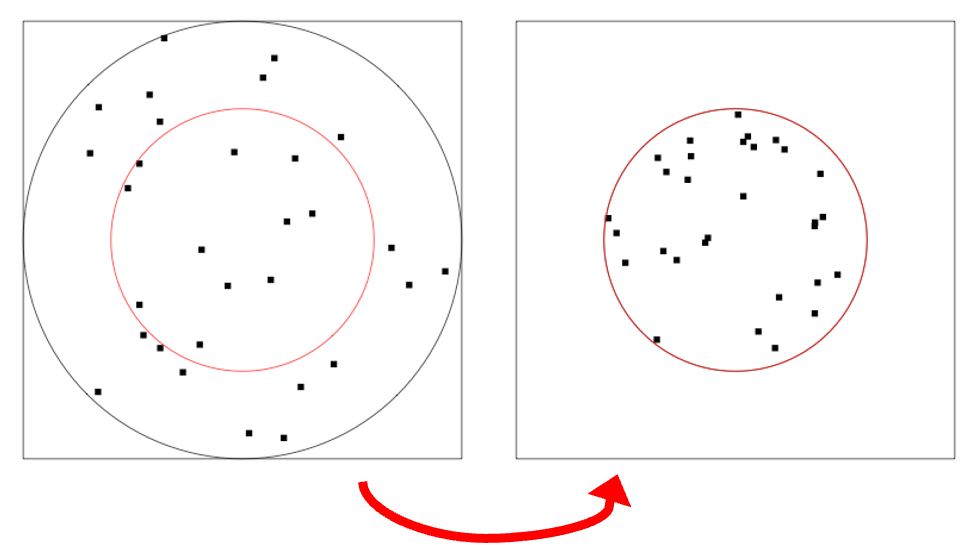}

\caption{Illustration of the inner step of a rare event ABC-MCMC algorithm
for a single point proposed in $\theta$-space. The black dots represent
draws from the model ($u$-points transformed by $H$ so that they
are in the data space). A uniform ABC kernel is used, so that the
circles represent regions of non-zero density for this kernel. The
outer circle in the left panel represents the ABC kernel for tolerance
$\epsilon_{t-1}$, with the black dots in the left panel representing
the $N_{u}$ points from the $\left(t-1\right)$th step of the algorithm.
The right panel shows the same points after reweighting, resampling
and an MCMC move. \label{fig:Illustation-of-the}}

\end{figure}

An illustration of the rare event approach can be found in figure
\ref{fig:Illustation-of-the}. 

\begin{algorithm}
\caption{Rare event SMC}\label{alg:rare-event-smc}
 
Simulate $N_{u}$ points, $\left\{ u_0^n \right\}_{n=1}^{N_{u}} \sim \phi\left( u \mid \theta \right)$ and set each weight $w_0^n = 1/N_u$;\

\For {$t=1:T$}
{
    \For(\tcp*[h]{reweight}) {$n=1:N_u$}
    {
        \eIf {$t=1$}
        {
            \[
            \tilde{w}^n_{t} = w^n_{t-1} P_{\epsilon_{t}}\left(y\mid H\left(u^n_{t-1},\theta\right)\right);
            \]
        }
        {
            \[
            \tilde{w}^n_{t} = w^n_{t-1} \frac{P_{\epsilon_{t}}\left(y\mid H\left(u^n_{t-1},\theta\right)\right)}{P_{\epsilon_{t-1}}\left(y\mid H\left(u^n_{t-1},\theta\right)\right)};
            \]
        }
    }
    
    $\left\{ w^n_{t} \right\}_{n=1}^{N_{u}} \leftarrow \mbox{ normalise}\left( \left\{ \tilde{w}^n_{t} \right\}_{n=1}^{N_{u}} \right)$;
    
    \If(\tcp*[h]{resample}) {some degeneracy condition is met}
    {
        \For{$n=1:N_u$}
        {
            Simulate the index $a^n_{t-1} \sim \mathcal{M}\left( \left( w_t^1, ..., w_t^{N_u} \right) \right)$ of the ancestor of particle $n$;
        }
        $w^n_{t} = 1/N_u$ for $n=1:N_u$;
    }
    
    \For(\tcp*[h]{move}) {$n=1:N_u$}
    {
        Simulate $u^n_{t} \sim K_t \left( \cdot \mid u^{a^n_{t-1}}_{t-1} \right)$, where $K_t$ is an MCMC move with invariant distribution  $P_{\epsilon_{t}}\left(y\mid H\left(u,\theta\right)\right)\phi\left( u \mid \theta \right)$ on $u$.
    }
}

\end{algorithm}

For this method to be efficient, the MCMC needs to be well-designed,
accounting for the fact that $\epsilon_{t}$ is decreasing at each
iteration. \citet{Prangle2016} find a slice sampler to be an efficient
choice here.

After running this algorithm, the ABC likelihood can be estimated using
\begin{equation} \label{eq:rare_event_lld}
\overline{l}\left(y\mid\theta\right) = \prod_{t=1}^T \sum_{n=1}^{N_{u}} \tilde{w}^n_{t}.
\end{equation}

\subsection{Overview of the rest of the paper}

This paper introduces a new approach to using ABC when $y$ is high-dimensional,
through combining ABC-SMC for exploring $\theta$-space with rare
event SMC for estimating the ABC likelihood. The aim is to gain the
strengths of both approaches.
\begin{itemize}
\item \textbf{Rare event ABC-MCMC} scales better than standard ABC-MCMC as the dimension
of the data grows. However, both methods are highly dependent on tuning
- primarily, how a user should choose $\epsilon$ to tradeoff bias and variance
expense in a sensible way. In addition, neither method can be used
to estimate the model evidence.
\item \textbf{ABC-SMC} allows $\epsilon$ to be chosen adaptively, and provides
an estimate of the model evidence. Also, it uses a population to explore
$\theta$-space, and can be more suited than MCMC for exploring multi-modal
targets. However, it uses the standard ABC likelihood estimate (equation (\ref{eq:abc_llhd_est_multiple})), which
leads to poor performance for high-dimensional $y$.
\end{itemize}
In section \ref{sec:Rare-event-ABC-SMC} we introduce the new method,
followed by presenting empirical results in section \ref{sec:Empirical-results}
and conclusions in section \ref{sec:Conclusions}.

\section{Rare-event ABC-SMC$^{2}$}\label{sec:Rare-event-ABC-SMC}

We begin this section by describing ABC-SMC, before moving on to the
new approach.

\subsection{ABC-SMC}

The use of SMC samplers in the ABC setting was pioneered by \citet{Sisson2007f}, with
the key idea being to use as a sequence of distributions a sequence
of ABC posteriors with decreasing tolerances $\epsilon_{t}$ for $t=1:T$,
where $\epsilon_{1}>...>\epsilon_{T}$. The first tolerance $\epsilon_{1}$
is typically chosen to give an ABC posterior close to the prior $p\left(\theta\right)$,
which is typically used as the initial distribution $\pi_{0}$. The
final tolerance $\epsilon_{T}$ is chosen to be the desired tolerance,
such as we would use in an ABC-MCMC algorithm.

\begin{figure}
\centering
\includegraphics[scale=0.45]{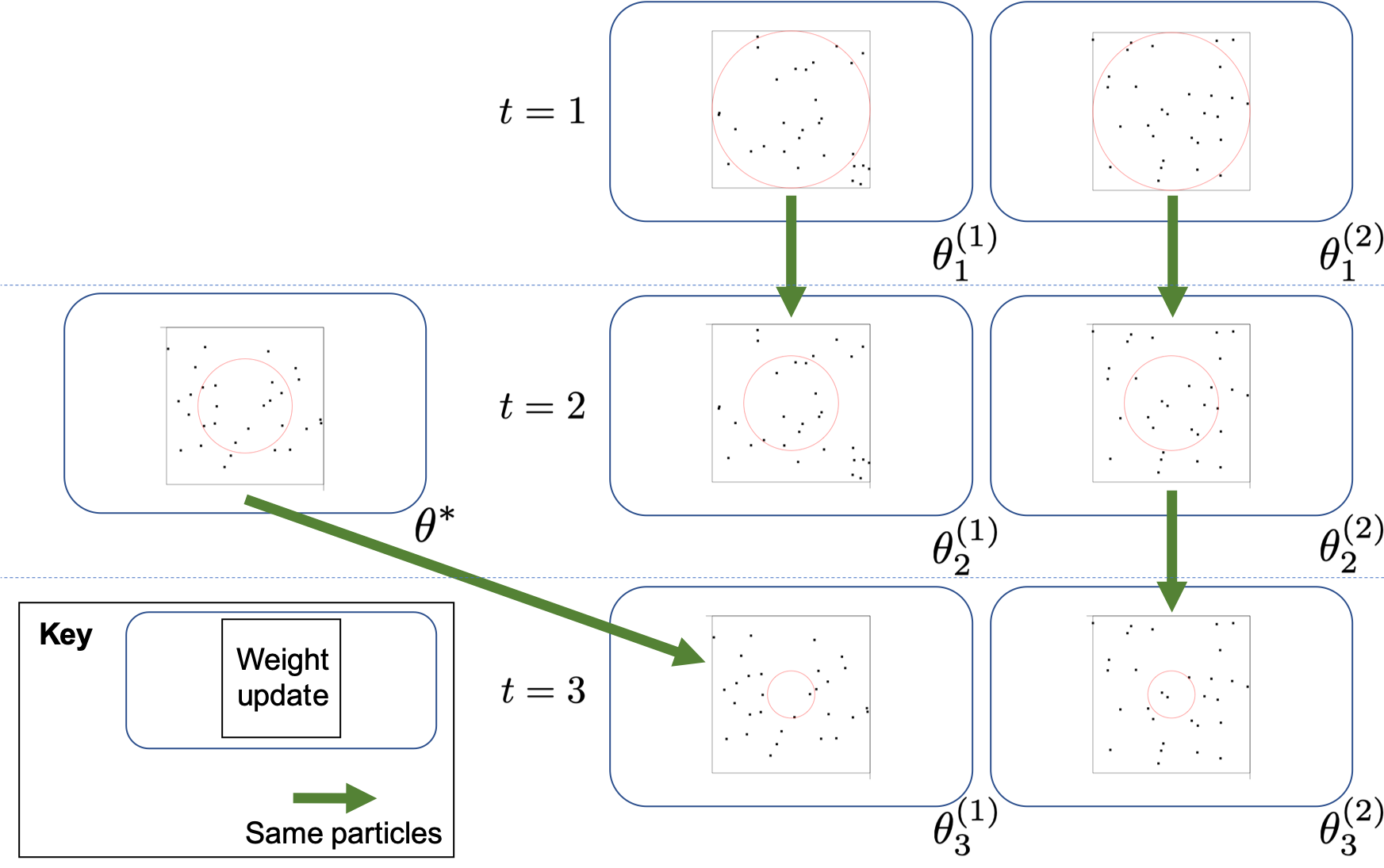}\caption{An illustration of an ABC-SMC algorithm with two $\theta$-particles
for three iterations, with one particle being updated using an ABC-MCMC
move. Each particle has its own draws from the model, which are represented
by black dots. A uniform ABC kernel is used, with the region of non-zero
density for this kernel being represented by the circle.Note that
once the draws from the model (the black dots) are made for each particle,
they remain the same throughout the algorithm, unless a $\theta$-particle
is replaced using an ABC-MCMC move (the proposed point for one particle being represented on the left of the figure), in which case it brings with it
its own draws from the model, which are made during the MCMC proposal.\label{fig:Illustration-of-ABC-SMC}}

\end{figure}

In this paper we use the variation on the idea introduced by \citet{DelMoral2012g}.
In this method the sequence of (unnormalised) targets is $P_{\epsilon_{t}}\left(y\mid x\right)f_{\theta}\left(x\right)p\left(\theta\right)$
for $t=1:T$, with the initial distribution being $f_{\theta}\left(x\right)p\left(\theta\right)$. In algorithm \ref{alg:abc-smc} we describe the form of ABC-SMC that uses a
likelihood estimate based on $N_{x}$ points drawn from the likelihood
for each $\theta$: for the $n$th one of these points drawn for the
$m$th particle in $\theta$-space, we use the notation $x_{t,\theta}^{n,m}$. Each step of the algorithm uses the likelihood estimate from equation (\ref{eq:abc_llhd_est_multiple}), i.e. for any $s,t$ we use
\begin{equation} \label{eq:estimated_abc_likelihood}
\hat{l}_s \left(y\mid\theta^n_t\right)=\frac{1}{N_{x}}\sum_{n=1}^{N_{x}}P_{\epsilon_s}\left(y\mid x_t^{n,m}\right).
\end{equation}
This version of the algorithm differs from the one in \citet{DelMoral2012g},
since we only use an MCMC move when a degeneracy condition is met.
The sampling from the mixture distribution in the algorithm is one
way of writing a resampling step, followed by an MCMC move. The reason
for writing the algorithm in this way is to make a direct comparison
with the SMC$^{2}$ algorithm of \citet{Chopin2013}, which we build
on in the next section. A figure illustrating the algorithm is shown
in figure \ref{fig:Illustration-of-ABC-SMC} and a video illustration
the steps of the algorithm can be found \href{https://youtu.be/CW4jQQUUnsI}{here}.

\begin{algorithm}
\caption{ABC-SMC}\label{alg:abc-smc}
 
Simulate $N_{\theta}$ points, $\left\{ \theta_{0}^{m} \right\}_{m=1}^{N_{\theta}} \sim p$ and set each weight $\omega_0^m = 1/N_{\theta}$;\

\For{$m=1:N_{\theta}$}
{
    \For{$n=1:N_{x}$}
    {
        $x_{0}^{n,m} \sim f\left( \cdot \mid \theta_0^m \right)$;\
    }
     
}

\For {$t=1:T$}
{

    \For(\tcp*[h]{reweight}) {$m=1:N_{\theta}$}
    {
        \eIf {$t=1$}
        {
            \[
            \tilde{\omega}^m_{t} = \omega^m_{t-1} \hat{l}_{1}\left(y\mid\theta_{t-1}^{m}\right);
            \]
        }
        {
        \[
            \tilde{\omega}^m_{t} = \omega^m_{t-1} \frac{\hat{l}_{t}\left(y\mid\theta_{t-1}^{m}\right)}{\hat{l}_{t-1}\left(y\mid\theta_{t-1}^{m}\right)};
            \]
        }
    }
    
    $\left\{ \omega^m_{t} \right\}_{m=1}^{N_{\theta}} \leftarrow \mbox{ normalise}\left( \left\{ \tilde{\omega}^m_{t} \right\}_{m=1}^{N_{\theta}} \right);$
    
    \If(\tcp*[h]{resample and move}) {some degeneracy condition is met}
    {
        \For{$m=1:N_{\theta}$}
        {
            Simulate $\left(\theta_t^{m}, x_{t}^{1:N_{x},m}\right)$ from the mixture distribution
            \[
            \sum_{i=1}^{N_{\theta}} \omega_t^{i} K_{t}\left\{ \cdot \mid \left(\theta_{t-1}^{i},x_{t-1}^{1:N_{x},i}\right) \right\},
            \]
            where $K_{t}$ is an ABC-MCMC kernel with respect to target $t$ in the SMC (lines 4-7 of algorithm \ref{alg:abc-mcmc} using tolerance $\epsilon_t$).
        }
        $\omega^m_{t} = 1/N_{\theta}$ for $m=1:N_{\theta}$;
        
    }
}

\end{algorithm}

\subsection{Rare-event ABC-SMC$^{2}$}

In this section, we introduce our new approach: the use of the structure
of the SMC$^{2}$ algorithm of \citet{Chopin2013} in the ABC setting,
through using rare event ABC to estimate likelihood ratios when required.
SMC$^{2}$ is designed for a state space model setting: $y_{1:t}$
are noisy observations of a latent time series $x_{1:t}$. The generative
model for this situation is specified in two parts: $f_{\theta}\left(x_{1:t}\right)$,
which models the dynamics of the latent time series, and $g_{\theta}\left(y_{1:t}|x_{1:t}\right)$.
which models the distribution of the observations. SMC$^{2}$ may
be used to estimate the posterior distribution on both $\theta$ and
$x_{1:t}$. It is set up using an ``external'' SMC on $\theta$-space,
and an ``internal'' SMC on $x$-space conditional on $\theta$. Each iteration of the external SMC algorithm necessitates running an additional step of a separate internal SMC algorithm for each particle in the external SMC.

The internal SMC has target $f_{\theta}\left(x_{1:t}\right)g_{\theta}\left(y_{1:t}|x_{1:t}\right)$
at iteration $t$. The weights calculated for each internal particle when updating the internal SMC
at iteration $t+1$ can be used to estimate $p\left(y_{t+1}\mid y_{1:t},\theta\right)=p\left(y_{1:t+1}\mid\theta\right)/p\left(y_{1:t}\mid\theta\right)$ for each external particle, this being the term needed in the weight update in the external SMC.

\begin{figure}
\centering
\includegraphics[scale=0.45]{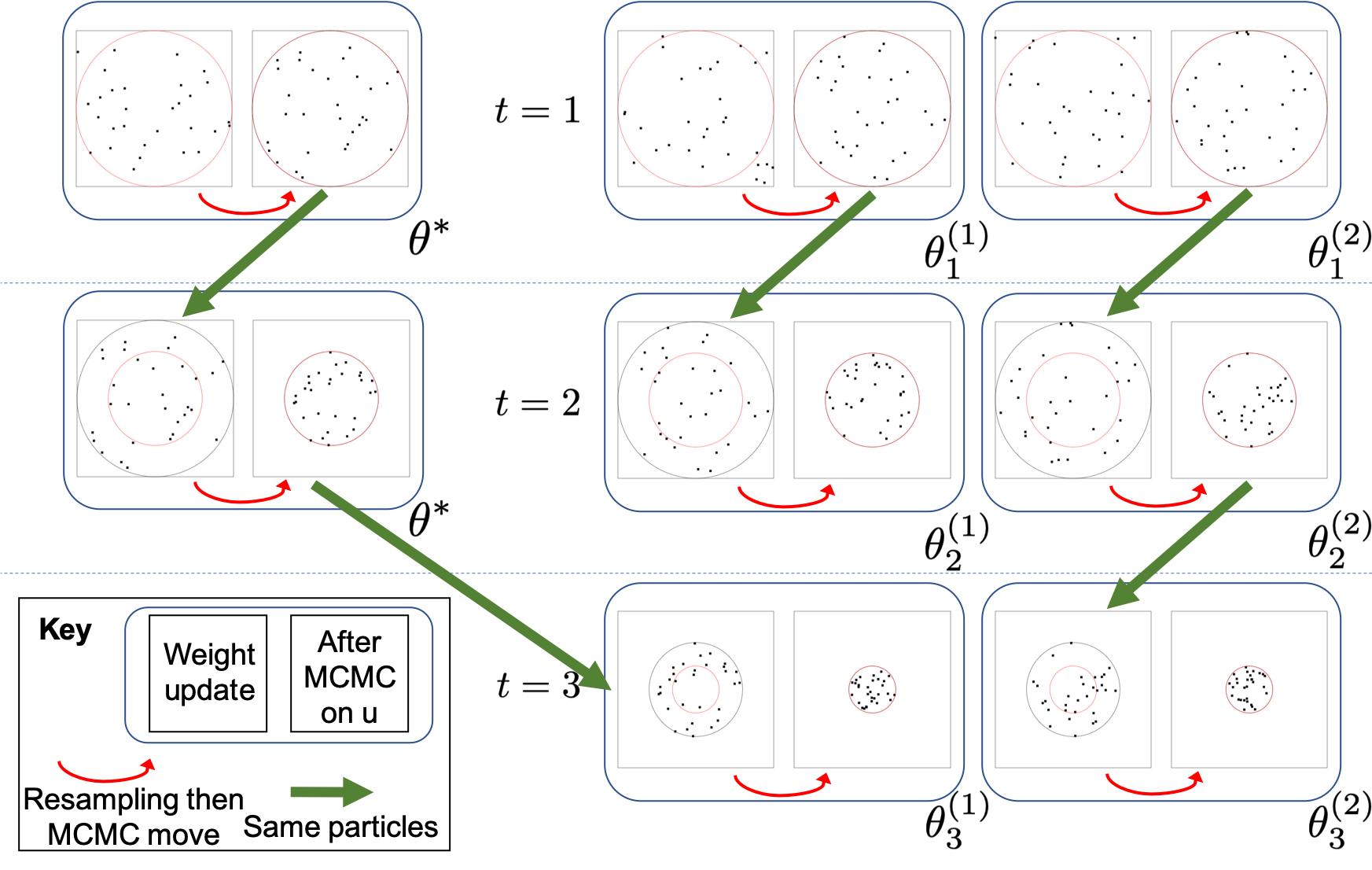}

\caption{An illustration of the ABC-SMC$^{2}$ algorithm with two
particles for three iterations, with one particle being updated using
a rare event ABC-MCMC move, as described in \citet{Prangle2016}.
Each particle has its own draws from the model, which are represented
by black dots. A uniform ABC kernel is used, with the region of non-zero
density for this kernel being represented by the red circle. Note
that, in contrast to ABC-SMC the draws from the model (the black dots)
are updated using MCMC moves at each iteration of the algorithm. When
a particle is replaced using an ABC-MCMC move, it must use this same
procedure of updates for its own draws from the model (as shown in the proposed particle on the left).\label{fig:An-illustration-of}}

\end{figure}

For our description of the algorithm, we follow as closely as possible,
the notation in the SMC$^{2}$ paper. The ``internal'' SMC algorithm is the rare event SMC method introduced
in the previous section. The
``external'' SMC algorithm is given in algorithm \ref{alg:abc-smc2}. A figure illustrating the algorithm is shown in figure \ref{fig:An-illustration-of} and a video illustration the steps of the algorithm can be found \href{https://youtu.be/vR1ZselG4ZY}{here}.

\begin{algorithm}
\caption{Rare event ABC-SMC$^{2}$ (RE-ABC-SMC$^2$)}
\label{alg:abc-smc2}

Simulate $N_{\theta}$ points $\left\{ \theta_0^{m} \right\}_{m=1}^{N_{\theta}} \sim p$ and set $\omega_0^{m} = 1/N_{\theta}$;\

\For{$m=1:N_{\theta}$}
{
    Simulate $N_{x}$ points $\left\{ u_0^{n,m} \right\}_{n=1}^{N_{x}} \sim \phi\left( \cdot \mid \theta_0^m \right)$;\
}

\For {$t=1:T$}
{

    \For(\tcp*[h]{reweight}) {$m = 1:N_{\theta}$}
    {

        \eIf {$t=1$}
        {
            Simulate $\left(u_{1}^{1: N_{u}, m}, a_{0}^{1: N_{u}, m}\right)$ using lines 3-16 of algorithm \ref{alg:rare-event-smc} when $t=1$, compute
            \[
            \widehat{l_{1}\left(y \mid \theta_0^{m}\right)}= \sum_{n=1}^{N_{u}} \tilde{w}^{n,m}_{1};
            \]
            \[
            \omega_1^{m} = \omega_0^{m} \widehat{l_{1}\left(y \mid \theta_0^{m}\right)};
            \]
        }
        {
            Simulate $\left(u_{t}^{1: N_{u}, m}, a_{t-1}^{1: N_{u}, m}\right)$ using lines 3-16 of algorithm \ref{alg:rare-event-smc}, compute:
            \[
            \widehat{\frac{l_{t}\left({y \mid \theta_{t-1}^{m}}\right)}{l_{t-1}\left(y \mid \theta_{t-1}^{m}\right)}}= \sum_{n=1}^{N_{u}} \tilde{w}^{n,m}_{t};
            \]
            \[
            \omega_t^{m} = \omega_{t-1}^{m} \widehat{\frac{l_{t}\left(y \mid \theta_{t-1}^{m}\right)} {l_{t-1}\left(y \mid \theta_{t-1}^{m}\right)} };
            \]

        }
	
    }

    $\left\{ \omega^m_{t} \right\}_{m=1}^{N_{\theta}} \leftarrow \mbox{ normalise}\left( \left\{ \tilde{\omega}^m_{t} \right\}_{m=1}^{N_{\theta}} \right);$
    
    \If(\tcp*[h]{resample and move}) {some degeneracy condition is met}
    {
        \For{$m=1:N_{\theta}$}
        {
            Simulate $\left(\theta_t^{m}, u_{1:t}^{1:N_{u},m}, a_{1:t-1}^{1:N_{u},m} \right)$ from the mixture distribution
            \[
            \sum_{i=1}^{N_{\theta}} \omega_t^{i} K_{t}\left\{ \cdot \mid \left(\theta_{t-1}^{i},u_{t}^{1:N_{u},i},a_{t}^{1:N_{u},i}\right) \right\},
            \]
            where $K_{t}$ is the MCMC move from \cite{Prangle2016}, i.e.:
            
            $i^* \sim \mathcal{M}\left( \left\{ \omega_t^{i} \right\}_{i=1}^{N_{\theta}} \right)$, then $\theta^* \sim q_t \left( \cdot \mid \theta_{t-1}^{i^*} \right)$, then run algorithm \ref{alg:rare-event-smc} up to $\epsilon_t$ conditional on $\theta^*$.
            
            Set $\theta_{t}^{m} = \theta^*$ and $u^{n,m}_{1:t}, a^{n,m}_{1:t-1}$ and $\tilde{w}^{n,m}_{1:t}$ to be the variables and unnormalised weights generated when running algorithm \ref{alg:rare-event-smc} with probability
            \[
            1\wedge\frac{p\left(\theta^{*}\right)}{p\left(\theta_{t-1}^{i^*}\right)}\frac{q\left(\theta_{t-1}^{i^*} \mid\theta^{*}\right)}{q\left(\theta^{*}\mid\theta_{t-1}^{i^*} \right)} \frac{\overline{l}\left(y\mid\theta^*\right)}{\prod_{t=1}^T \sum_{n=1}^{N_{u}} \tilde{w}^{n,^*}_{t}},
            \]
            where $\overline{l}$ is defined in equation \ref{eq:rare_event_lld};
            
            Else set $\theta_{t}^{m} = \theta_{t-1}^{i^*}$, $\tilde{w}^{n,m}_{1:t} = \tilde{w}^{n,i^*}_{1:t}$, $u^{n,m}_{1:t}=u^{n,i^*}_{1:t}$ and $a^{n,m}_{1:t-1}=a^{n,i^*}_{1:t-1}$.
        }
        $\omega^m_{t} = 1/N_{\theta}$ for $m=1:N_{\theta}$;
        
    }
    
}
\end{algorithm}

\subsection{Implementation details} \label{sec:implementation}

Algorithm \ref{alg:rare-event-smc} may be modified in several ways in order to, at each iteration, make use of the current set of weighted particles to inform subsequent steps of the algorithm. These modifications allow many of the tuning parameters of the algorithm to be chosen adaptively. This section describes the adaptive approaches used in this paper.

\subsubsection{Adapting the sequence of tolerances} \label{sec:adapt_tol}

One of the appealing properties of ABC-SMC (algorithm \ref{alg:abc-smc}) in comparison with ABC-MCMC (algorithm \ref{alg:abc-mcmc}) is that the choice of the tolerance $\epsilon$ may be automated in ABC-SMC. The most commonly-used approach is that of \cite{DelMoral2012g}. This method adds an additional routine before line 8 of algorithm \ref{alg:abc-smc} to determine the choice of $\epsilon_t$ at to be used in the current iteration of the SMC. We make use of the fact that $\hat{l}_{t}\left(y\mid\theta_{t-1}^{m}\right)$ can be computed very cheaply for different values of $\epsilon$ given that $\left\{ x_t^{n,m} \right\}_{n=1}^{\theta_x}$ have already been simulated. This means that the weights $\left\{ \tilde{\omega}^m_{t} \right\}_{m=1}^{N_{\theta}}$ may be calculated for many different choices of $\epsilon$, with the most appropriate value of $\epsilon$ chosen. Usually a bisection algorithm is used to search for the most appropriate $\epsilon$ within the range $[0,\epsilon_{t-1}]$. The most appropriate $\epsilon$ is usually chosen by finding the $\epsilon$ that results in the estimated conditional effective sample size (CESS) \citep{Zhou2015} being closest to some proportion $\beta \in (0,1)$ of $N_{\theta}$. The CESS, defined by

\[
{CESS}_{t}=\frac{N_{\theta} \left(\sum_{m=1}^{N_{\theta}} w_{t-1}^{m} \frac{\hat{l}_{t}\left(y\mid\theta_{t-1}^{m}\right)}{\hat{l}_{t-1}\left(y\mid\theta_{t-1}^{m}\right)}\right)^{2}}{\sum_{m=1}^{N_{\theta}} w_{t-1}^{m}\left(\frac{\hat{l}_{t}\left(y\mid\theta_{t-1}^{m}\right)}{\hat{l}_{t-1}\left(y\mid\theta_{t-1}^{m}\right)}\right)^{2}},
\]
was introduced in as more appropriate measure of degeneracy between two successive steps of the algorithm than the ESS in the case when resampling is not performed at every step. This approach ensures that the sequence of distributions is not chosen to change so quickly that the SMC becomes degenerate.

A similar approach may be used in algorithm \ref{alg:rare-event-smc}. At the same stage of the algorithm (before line 6 in algorithm \ref{alg:rare-event-smc}), a bisection routine may be added to find $\epsilon_{t}$ for the next step of the algorithm. Again the CESS may be used to chose the most appropriate $\epsilon_{t}$, in this case given by

\[
{CESS}_{t}=\frac{N_{\theta} \left(\sum_{m=1}^{N_{\theta}} w_{t-1}^{m} \widehat{\frac{l_{t}\left(y \mid \theta_{t-1}^{m}\right)} {l_{t-1}\left(y \mid \theta_{t-1}^{m}\right)} }\right)^{2}}{\sum_{m=1}^{N_{\theta}} w_{t-1}^{m}\left(\widehat{\frac{l_{t}\left(y \mid \theta_{t-1}^{m}\right)} {l_{t-1}\left(y \mid \theta_{t-1}^{m}\right)} }\right)^{2}},
\]
Once more the calculation of this criterion is computationally cheap given the variables $\left\{ u^{n,m}_{1:t} \right\}_{n=1}^{\theta_u}$ generated at the previous step of the algorithm. In this case, to calculate the ratio $\widehat{\frac{l_{t}\left(y \mid \theta_{t-1}^{m}\right)} {l_{t-1}\left(y \mid \theta_{t-1}^{m}\right)} }$ for each candidate $\epsilon$, we need only to run lines 3-9 of algorithm \ref{alg:rare-event-smc}.

\subsubsection{Adapting the MCMC proposals}

Algorithm \ref{alg:abc-smc2} makes use of two MCMC steps: one in the external SMC, which we may think of as a move on $\theta$, and the other in the internal SMC, which is a move on $u$. \citet{Prangle2016} noted the importance of the internal MCMC move adapting to match the changing scale of its target as $\epsilon_t$ changes, and the same is true of the external MCMC move. \citet{Prangle2016} used a slice sampler to achieve this effect, but other approaches are also possible, such as using the current population of particles to estimate the scale of a proposal for a Metropolis-Hastings algorithm, as is used for example in \citet{everitt_delayed_2021}.

\subsubsection{Adapting the number of MCMC moves} \label{sec:adapting-number-mcmc}

As ABC-SMC, algorithm \ref{alg:abc-smc} runs, $\epsilon_t$ decreases, and it becomes more unlikely that the simulation $x$ is close to $y$. The result is that the acceptance rate of the ABC-MCMC moves used by the algorithm decreases as $t$ increases. Eventually the acceptance rate drops to zero, and whilst the tolerance $\epsilon_t$ continues to decrease and thus resampling is performed, the population ends up consisting of many duplicated particles. We observed a similar effect when running RE-ABC-SMC$^2$, but for a much lower tolerance than for ABC-SMC.

One approach to avoiding this issue is to adapt the number of MCMC iterations used as the algorithm progresses. \citet{south_sequential_2019} suggest the following approach to choosing the number of iterations for which to run a Metropolis-Hastings algorithm: at each SMC iteration they examine the acceptance rate (denoted $\hat{p}_{\text {acc}}$) across all particles of the first iteration of the MCMC move, and choose the number of iterations to be
\[
\left\lceil\frac{\log (c)}{\log \left(1-\hat{p}_{\mathrm{acc}}\right)}\right\rceil,
\]
with $\left\lceil\cdot\right \rceil$ denoting the ceiling function, such that there is an estimated probability of $1-c$ that each particle is moved at least once.

In the following section we use this adaptive approach in each implementation of ABC-SMC or RE-ABC-SMC$^2$, taking $c=0.2$.

\subsubsection{Splitting different sources of randomness} \label{sec:splitting}

Section \ref{sec:unpacking} describes how we make use of the reparameterisation of a simulator model as a deterministic transformation $H\left( u,\theta \right)$ of the parameter $\theta$ and the random vector $u$, where $u$ encompasses all of the stochasticity in the model. We then use MCMC moves on $u$ within the algorithm. However, for some models, such as that in section \ref{sec:random-graph}, the space of $u$ can be extremely complicated: e.g. it may be of variable dimension. Whilst in principle it is possible to design MCMC moves for such spaces, in practice it may not be possible to do so effectively. Therefore it may be desirable to split $u$ in two parts: $u_s$, which will be updated by the MCMC, and $u_r$ which will not be moved by the MCMC. This requires no change to algorithm \ref{alg:abc-smc2}. Although the MCMC move on $u$ is not irreversible, the algorithm is still a valid SMC sampler, as can be seen through noting that the ABC-SMC in algorithm \ref{alg:abc-smc} the a special case of algorithm \ref{alg:abc-smc2} where $u=u_r$, so that no MCMC moves are used on $u$.

\section{Empirical results\label{sec:Empirical-results}}

\subsection{Gaussian model}

This section considers a slightly amended version of a Gaussian model studied in \citet{Prangle2016}. We simulated $d$ points $y = \left\{y_i \right\}_{i=1}^d$ from a truncated univariate Gaussian distribution with $\mu=0$, $\sigma=3$, lower bound $a=0$ and no upper bound. The aim of the inference is to estimate the posterior distribution of $\theta = \sigma$ given $y$. We used the uniform prior $\theta \sim \mathcal{U}\left( 0, 10\right)$

$P_{\epsilon}\left( y \mid x \right)$ was chosen to be the uniform distribution $\mathcal{U} \left( (y-x)^2 - \epsilon, (y-x)^2 + \epsilon \right)$. We use $H\left(u,\theta \right) = \lVert \theta \Phi^{-1}\left(u\right) \rVert$ for $u\in [0,1]$, where $\Phi^{-1}$ is the inverse Gaussian cdf. The MCMC move used to update $u$ is the slice sampler from \citet{Prangle2016}. For the move step on $\theta$, we used a Metropolis-Hastings algorithm with a truncated (at zero) Gaussian proposal with variance chosen to be equal to the variance of the weighted particles after the reweighting step. The number of iterations of the MCMC was determined adaptively as described in section \ref{sec:adapting-number-mcmc}. In all SMC algorithms we resample when the ESS drops below the proportion $\alpha=0.5$ of the number of particles, and we chose the sequence of $\epsilon_t$ adaptively such that $\epsilon_t$ is as close as possible to giving a CESS of $\beta N_{\theta}$, where $\beta = 0.9$.

We ran ABC-SMC and RE-ABC-SMC$^2$ on three different scenarios: with $d=25$, $50$ and $100$. As described in section \ref{sec:unpacking}, we expect the performance of ABC-SMC to significantly deteriorate as $d$ increases, whilst the performance of RE-ABC-SMC$^2$ should not deteriorate as quickly. For each scenario, we initially ran a pilot run of RE-ABC-SMC$^2$, and terminated it when the acceptance rate of the Metropolis-Hastings moves became close to zero: this corresponded to a tolerance of $\epsilon=3$ for $d=25$ (using $N_{\theta}=250, N_u=100$), $\epsilon=5$ for $d=50$ (using $N_{\theta}=1000, N_u=500$) and $\epsilon=10$ for $d=100$ (using $N_{\theta}=1000, N_u=2000$). We recorded the runtime for RE-ABC-SMC$^2$ in each scenario, then ran ABC-SMC for the same runtime as was used by RE-ABC-SMC$^2$. ABC-SMC used $N_x = 1$ in every case, and for $N_{\theta}$, a number of particles that gave it approximately the same runtime per iteration as RE-ABC-SMC$^2$: $N_{\theta} =$ 3$\times 10^5, 7\times10^6,2.5\times10^7$ for $d=25,50,100$ respectively.

\begin{figure}
    \centering
    \subfloat{\includegraphics[width=0.42\textwidth]{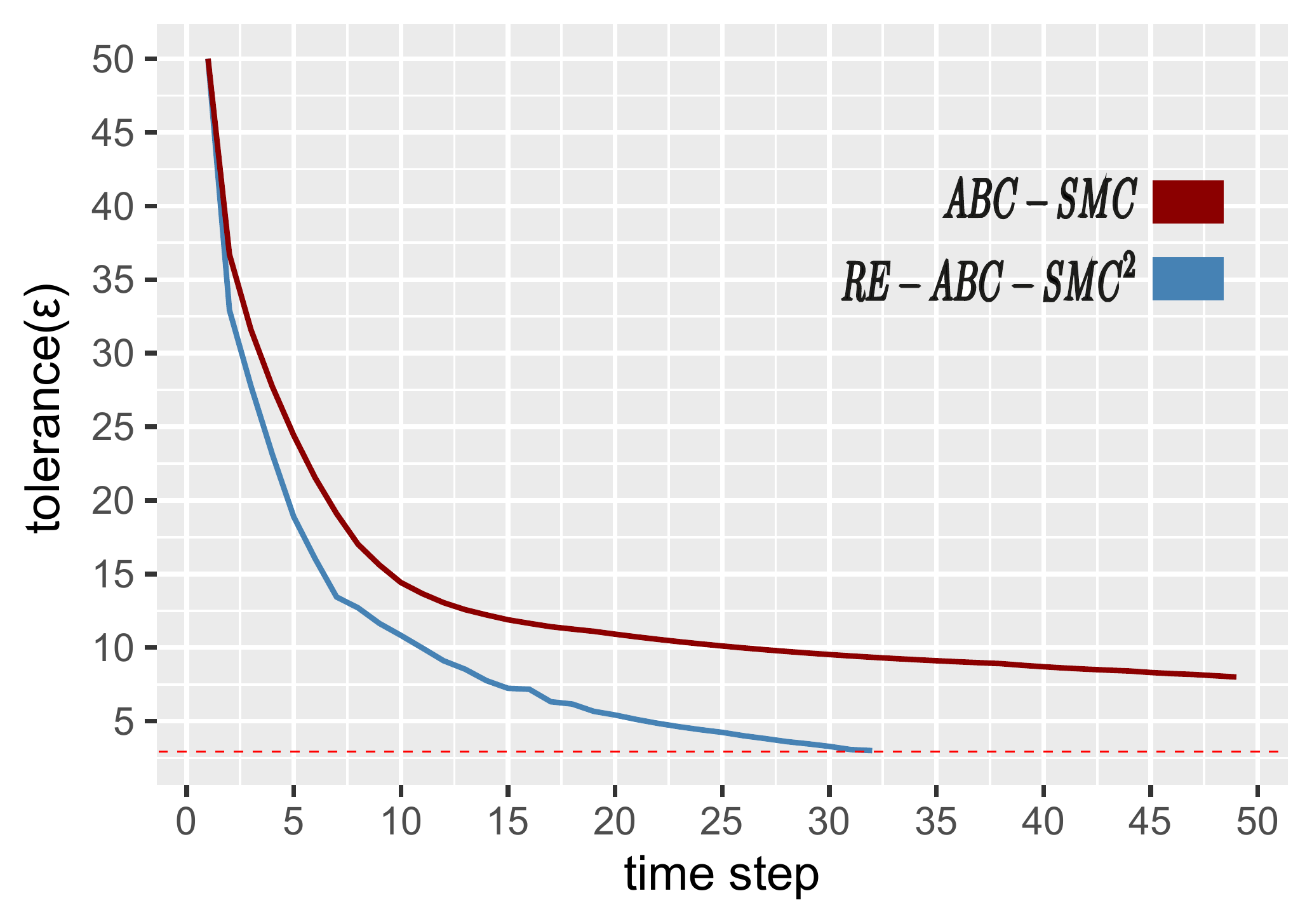}}%
    \qquad
    \subfloat{\includegraphics[width=0.42\textwidth]{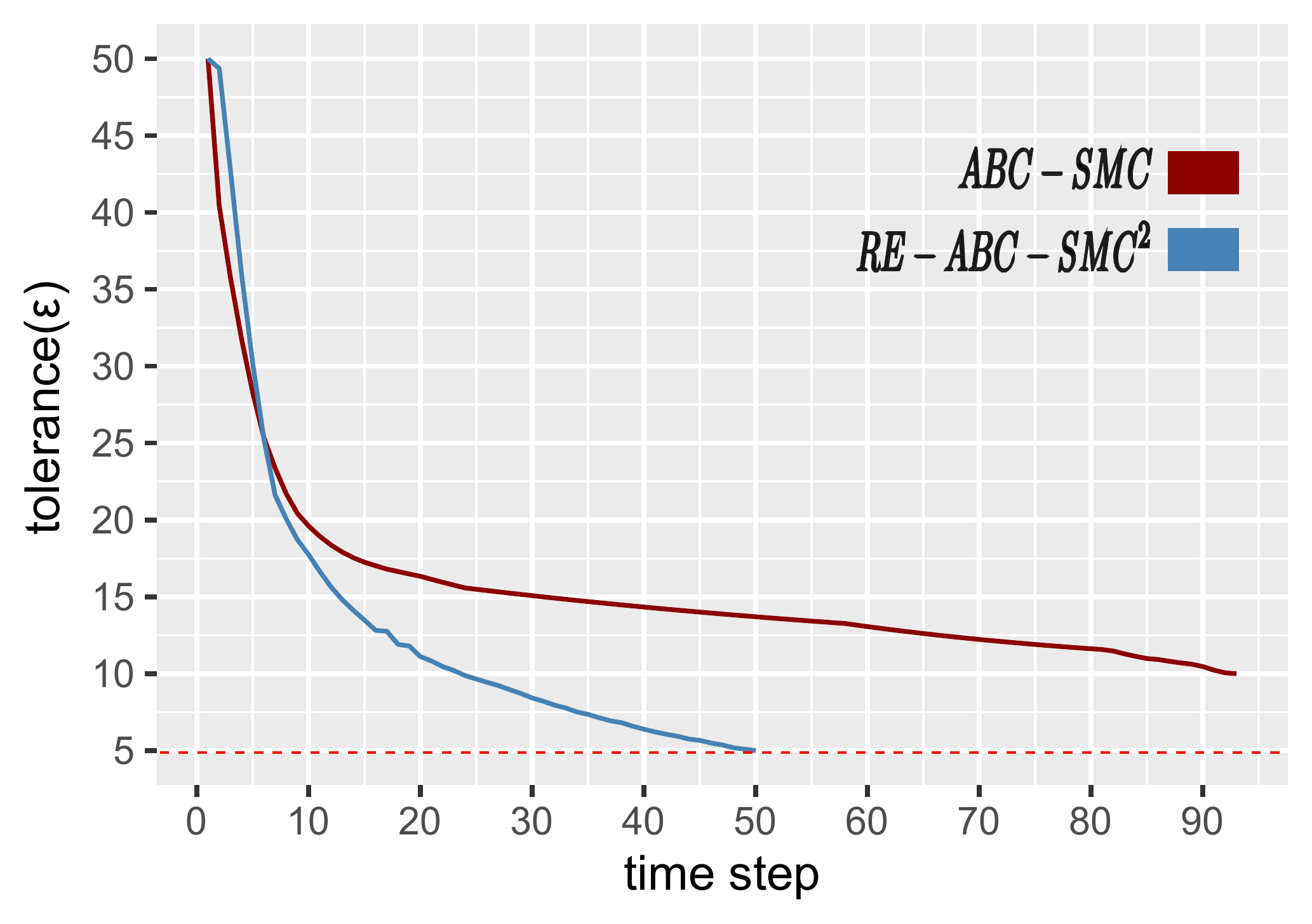} }\\
    \subfloat{\includegraphics[width=0.42\textwidth]{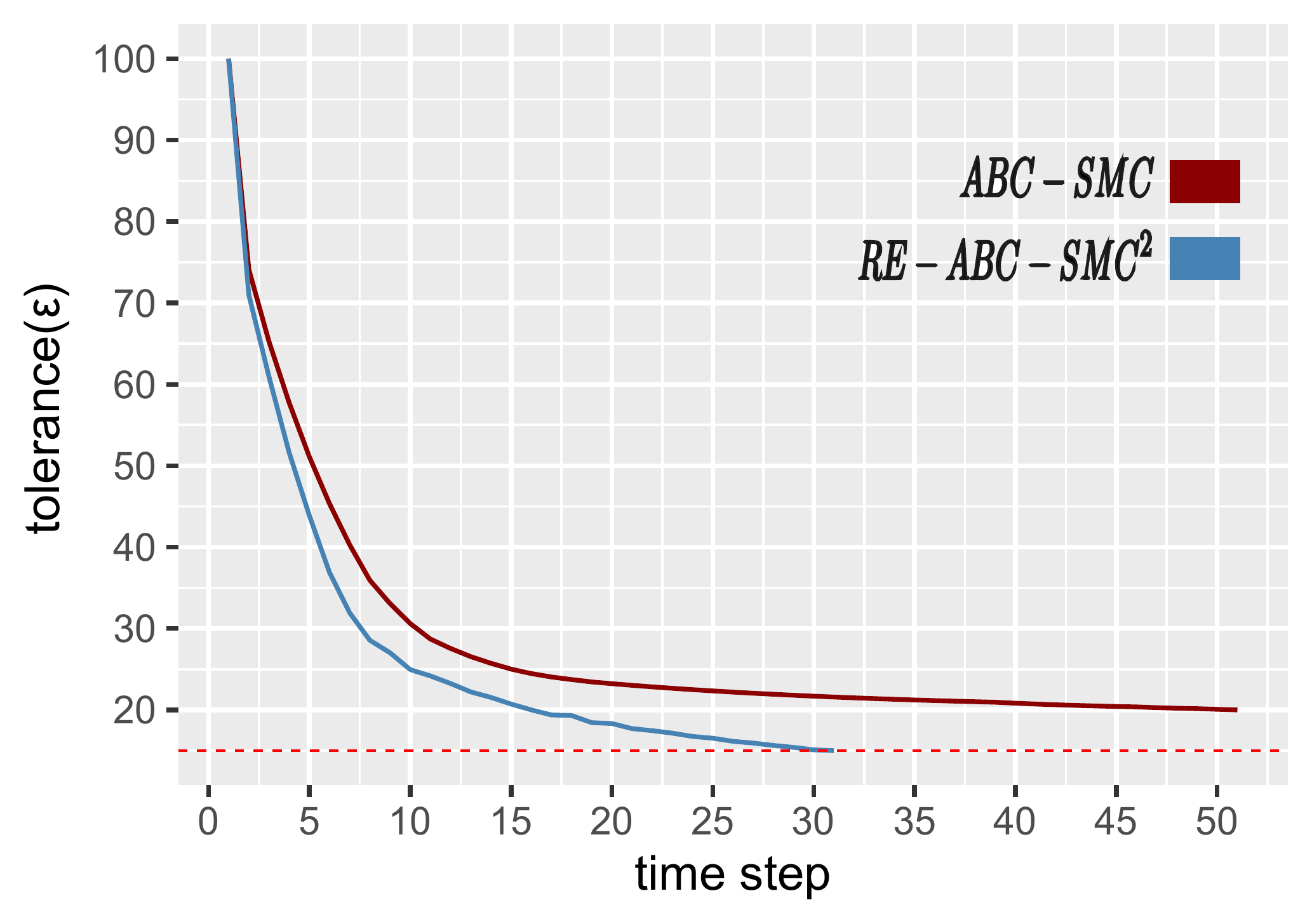} }%
    \caption{Comparison of adaptive schedules for ABC-SMC and RE-ABC SMC$^2$ in dimension $d=20$ (left, top), $d=50$ (right, top) and $d=100$ (bottom).}%
    \label{fig:resmc2gaus25050}%
\end{figure}

Figure \ref{fig:resmc2gaus25050} shows the evolution of $\epsilon_t$, as found using the adaptive algorithm from section \ref{sec:adapt_tol} over SMC iterations. We observe that the RE-ABC-SMC$^2$ algorithm is able to achieve a much lower tolerance than ABC-SMC, and that this tolerance decreases more quickly.

\begin{figure}
    \centering
    \includegraphics[width=0.75\textwidth]{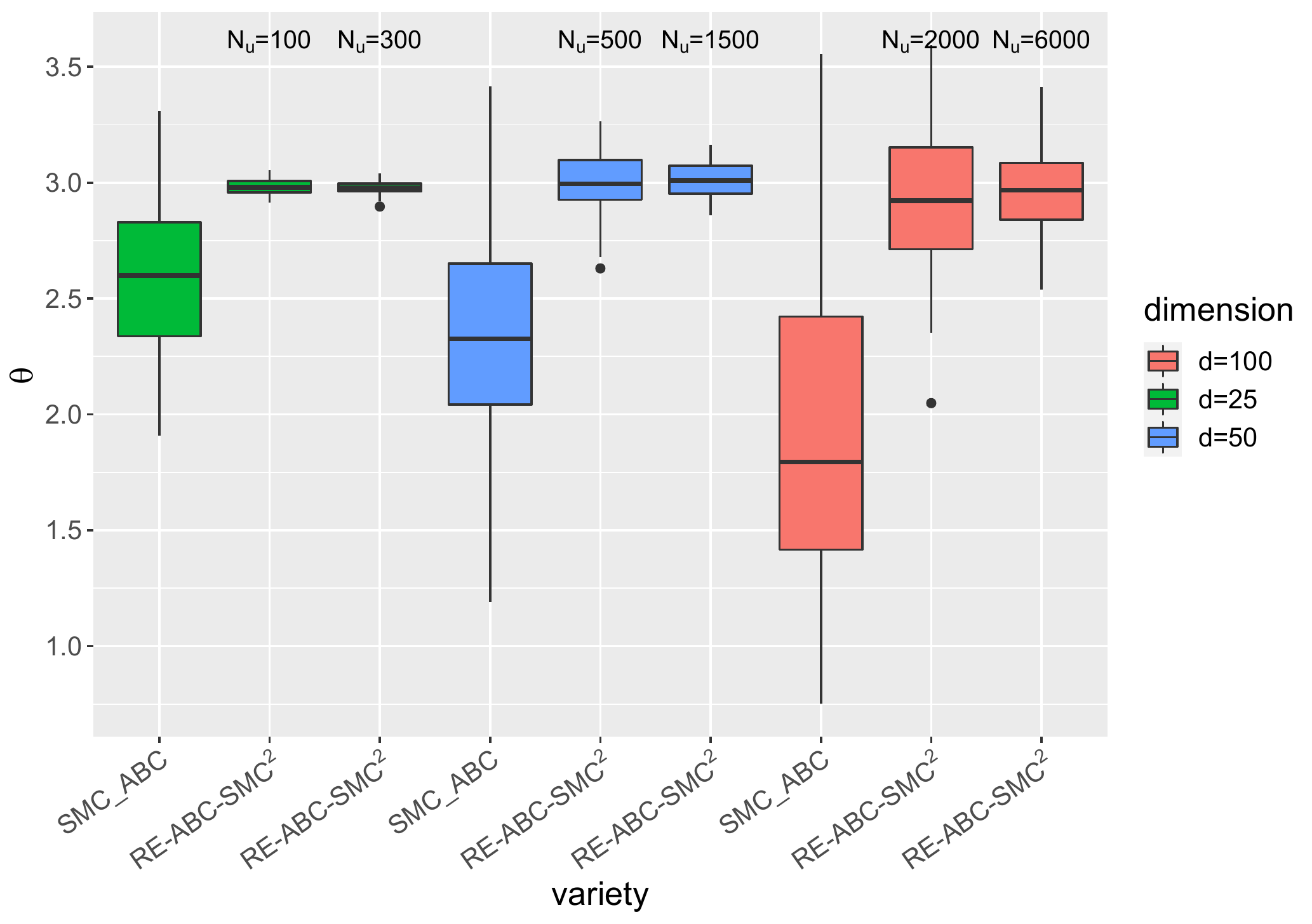}
    \caption{Box plots of the estimated posterior means estimated by ABC-SMC and RE-ABC SMC$^2$ over 50 runs, for the three different scenarios: left $d=25$; middle $d=50$ and right $d=100$.}
    \label{fig:resmc2gaussvariances}
\end{figure}

A lower tolerance does not necessarily mean that an algorithm provides better estimates of the posterior. We also examined the estimates of the posterior mean from each algorithm at the point at which it was terminated. We ran each algorithm 50 times: box plots of the posterior mean estimates are shown in figure \ref{fig:resmc2gaussvariances}. The left plot in each scenario corresponds to ABC-SMC, and the middle corresponds to RE-ABC-SMC$^2$ with the parameters described in the previous paragraph. Recall that the true value of $\theta$ is 3. We observe that in all cases, the estimated posterior mean from RE-ABC-SMC$^2$ is centred around the true value with comparatively lower variance than that from ABC-SMC. We also observe that the estimate from ABC-SMC is biased: evidence that the tolerance achieved by ABC-SMC is not low enough to produce accurate results. Further, we ran RE-ABC-SMC$^2$ with an increased number of $N_u$ particles (the box plots on the right of each scenario) and observed an improved performance, as expected.


\subsection{Parameter inference in duplication-divergence random graph models} \label{sec:random-graph}

Various random graph models have been proposed to model the formation
of complex networks in biology. In particular in the study of protein
interaction networks, growth models where networks are built through
the addition of nodes over many steps, have been considered, such
as linear preferential attachment models \citep{barabasiEmergenceScalingRandom1999a},
producing scale-free networks, and biologically inspired node duplication
models \citep{vazquezModelingProteinInteraction2003,pastor-satorrasEvolvingProteinInteraction2003a}.
It has been shown that the final structure of some of these models
can depend heavily on the unobserved initial state of the network
used in the growth model \citep{hormozdiariNotAllScaleFree2007}.

In this example we consider a duplication-divergence model of network
growth \citep{vazquezModelingProteinInteraction2003,pastor-satorrasEvolvingProteinInteraction2003a},
where starting from an initial seed network, the network is grown
in discrete steps, selecting a node at random to duplicate at each
step. The duplicated node retains the edges of the original with a
probability $p$, and forms a link to the original node with probability
$r$. This mimics the biological process of gene duplication, where
genes are copied and then diverge in function over time, with their
protein products losing some of the interactions of the original gene
in the process.

Our aim in this section is to estimate the parameters $p$ and $r$ given simulated network data $y$. To simulate $y$, we begin with a seed network composed of two cliques
of fully connected nodes, then generate connections between them formed uniformly out of the set of possible clique to clique connections with fixed probability. Seed networks of this kind were shown in \citet{hormozdiariNotAllScaleFree2007}
to produce networks with markedly different structural properties than other types of seed. From this seed, we then generate a random network with $d=100$ nodes using the duplication-divergence process with parameters $p=0.5$ and $r=0.2$.

The full data generation process used as the model within the ABC algorithms first simulates a seed network using an Erd\H{o}s-R\'{e}nyi random graph with $d_s=20$ nodes: the existence of each of the $d_s(d_s-1)/2$ edges is modelled by independent Bernoulli distributions with probability $a=0.3$. We denote the edges in the seed network by $u_s$, Then, conditional on the seed network, the duplication-divergence process is simulated using parameters $\theta = \left(p,r \right)$. The full details are: given an undirected seed network represented as a set of nodes $\mathcal{N}$ and edges $\mathcal{E}$, at each iteration of the model, we select a node $x_i\in \mathcal{N}$ uniformly at random to duplicate. To do so we first create a new node in the network $x^*$, with no edges. Then we take all nodes $x_j\mid(x_i,x_j)\in\mathcal{E}$ neighbouring $x_i$, and attach them to the new node $x^*$ forming new edges $(x_j,x^*)$, each with probability $p$. Finally the new node $x^*$ is connected to the node that was duplicated, $x_i$, forming an edge $(x_i,x^*)$, with probability $r$. This process is repeated until the desired number of nodes ($d=100$) in the network is reached. We denote the variables generated in the duplication-divergence process as $u_r$.

To apply the RE-ABC-SMC$^{2}$ methodology in this context, we use the idea in section \ref{sec:splitting} of dividing the $u$ variables into two parts: the $u_s$ variables will be updated using MCMC moves in the internal SMC, and the $u_r$ variables will not be updated. We make this choice since the dimension of $u_r$ will change dependant on its value in a complex way, and would be difficult to update effectively using MCMC. Specifically, some of the $u_r$ will correspond to the choice of whether or not to form an edge between a newly created node $x^*$, and one of the neighbours $x_j$ of the existing node $x_i$ chosen to be duplicated. However the number of such $u_r$ will depend on the number of edges $x_i$ has, which in turn could depend on previous values of $u_r$ used to make decisions on whether edges were formed when node $x_i$ was created.

We used the uniform prior $\mathcal{U}\left( 0, 1\right)$ for both $p$ and $r$. $P_{\epsilon}\left( y \mid x \right)$ was chosen to be the uniform distribution $\mathcal{U} \left( d(y,x) - \epsilon, d(y-x) + \epsilon \right)$, where $d(y,x)$ is an approximation to the edit distance between the two graphs $y$ and $x$. The edit distance is defined as the smallest number of edges that would need to either be added to or deleted from $y$ or $x$ for the two graphs to become isomorphic. This is prohibitively computationally expensive to calculate directly, but can be approximated as in \citet{thorneGraphSpectralAnalysis2012a} using the ordered eigenvalues $\alpha_1,\ldots,\alpha_d$ and $\beta_1,\ldots,\beta_d$ of the adjacency matrices of $y$ and $x$ respectively as
\[
    d(y,x) \approx \sum_i (\alpha_i-\beta_i)^2.
\]
To construct an MCMC kernel on $u^s$, we apply a Metropolis-Hastings sampler with a proposal that either adds or deletes an edge in the seed network with equal probability $q_{add}=q_{del}=0.5$. When an edge addition proposal is chosen, one of the $N(N-1)/2-|\mathcal{E}|$ possible pairs of unconnected nodes is selected uniformly at random, and an edge added between them. For an edge deletion proposal, one of the $|\mathcal{E}|$ edges in the seed network is chosen uniformly at random and deleted. 2 sweeps of the Gibbs sampler were used at every SMC iteration. For the move step on $\theta$, we used a Metropolis-Hastings algorithm with a truncated (from 0 to 1) Gaussian proposal with variance chosen to be equal to the variance of the weighted particles after the reweighting step. The number of iterations of the MCMC was again determined adaptively as described in section \ref{sec:adapting-number-mcmc}. In all SMC algorithms again we resample when the ESS drops below the proportion $\alpha=0.5$ of the number of particles, and we chose the sequence of $\epsilon_t$ adaptively such that $\epsilon_t$ is as close as possible to giving a CESS of $\beta N_{\theta}$, where $\beta = 0.9$.

\begin{figure}
    \centering
    \includegraphics[width=0.7\textwidth]{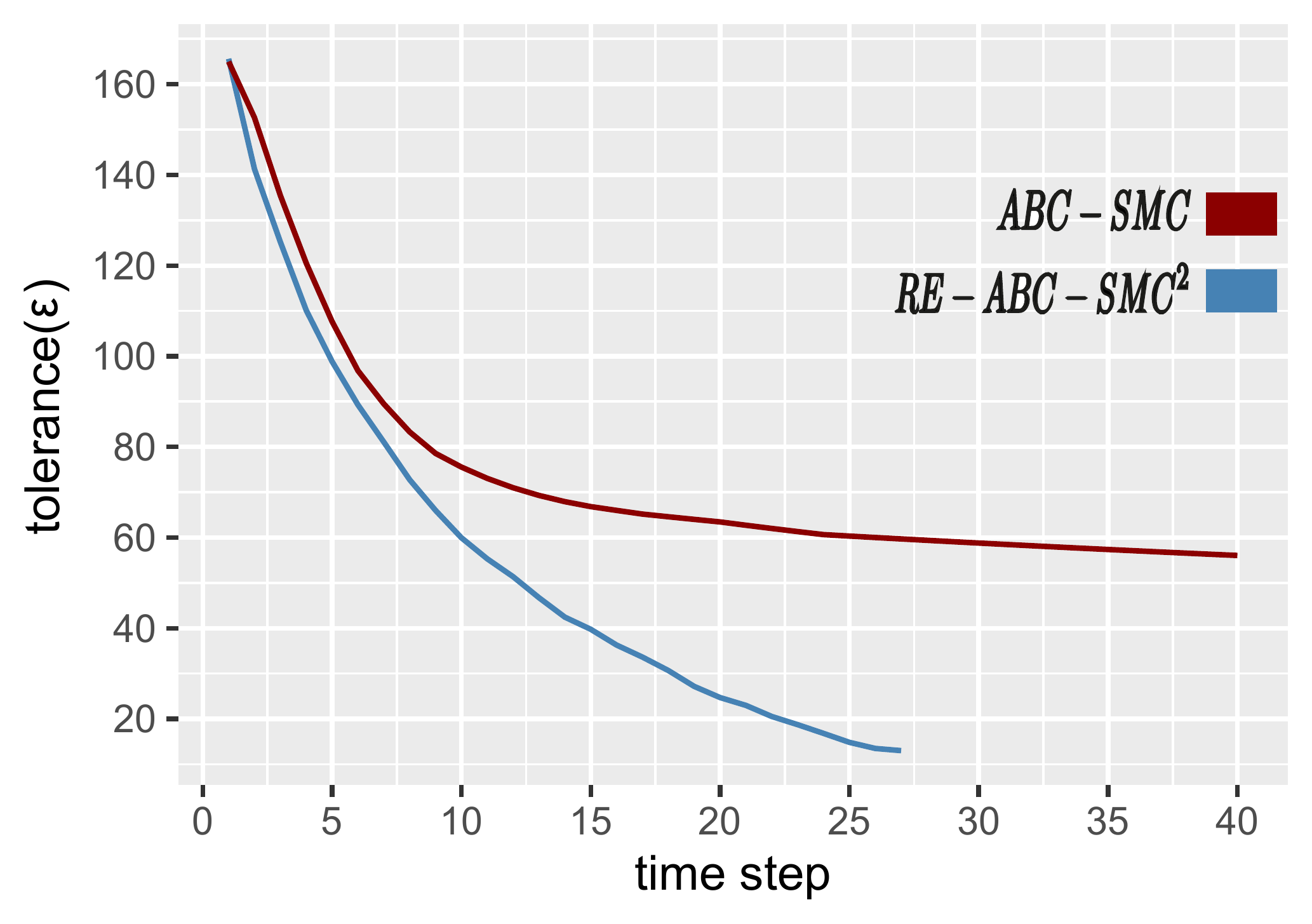}
    \caption{Comparison of adaptive schedules for ABC-SMC and RE-ABC SMC$^2$ on the duplication-divergence random graph model tolerance over time for ABC-SMC and RE-ABC-SMC$^2$.}
    \label{fig:resmc2networktol}
\end{figure}
\begin{figure}
    \centering
    \subfloat{{\includegraphics[width=0.4\textwidth]{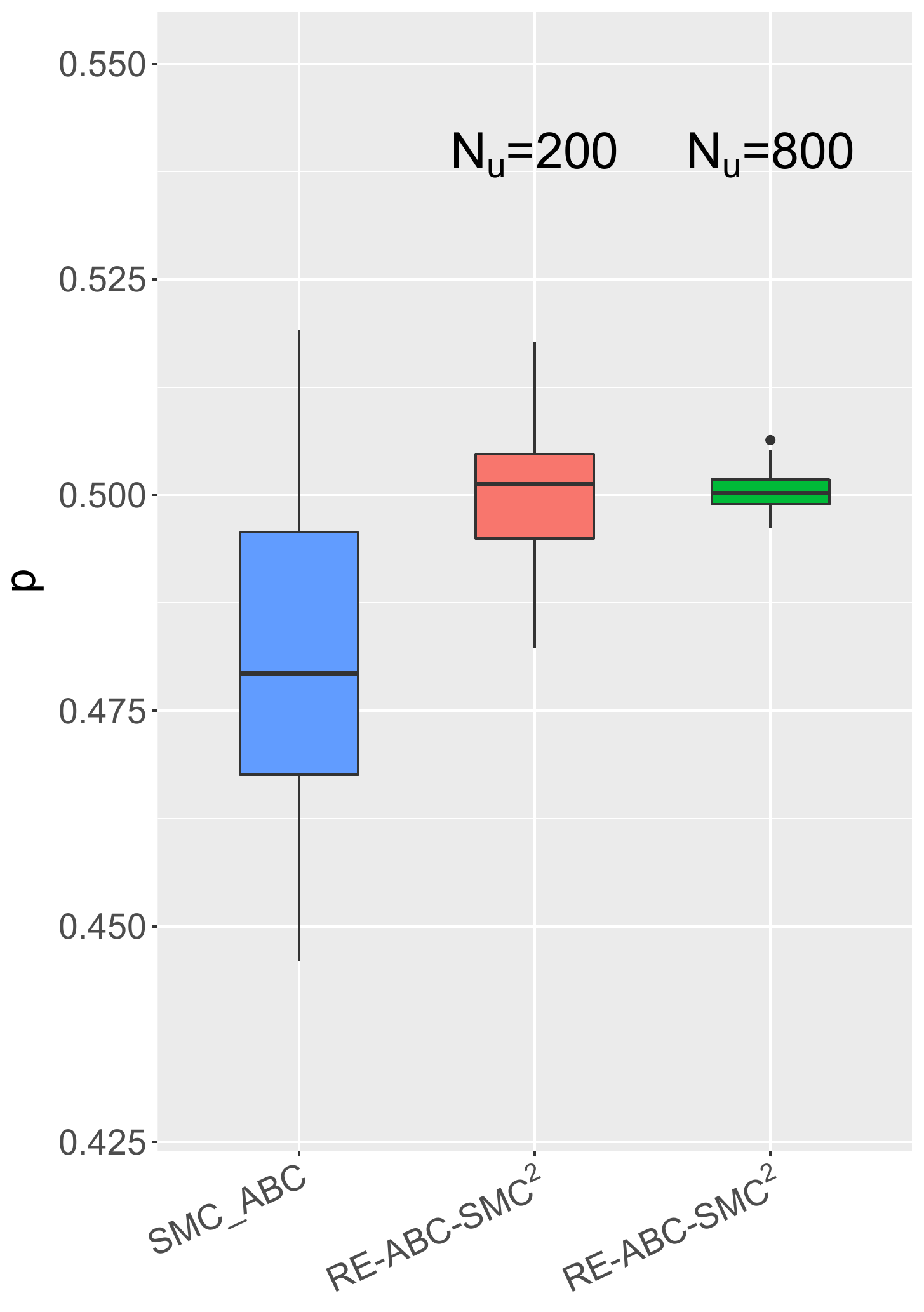} }}%
    \qquad
    \subfloat{{\includegraphics[width=0.4\textwidth]{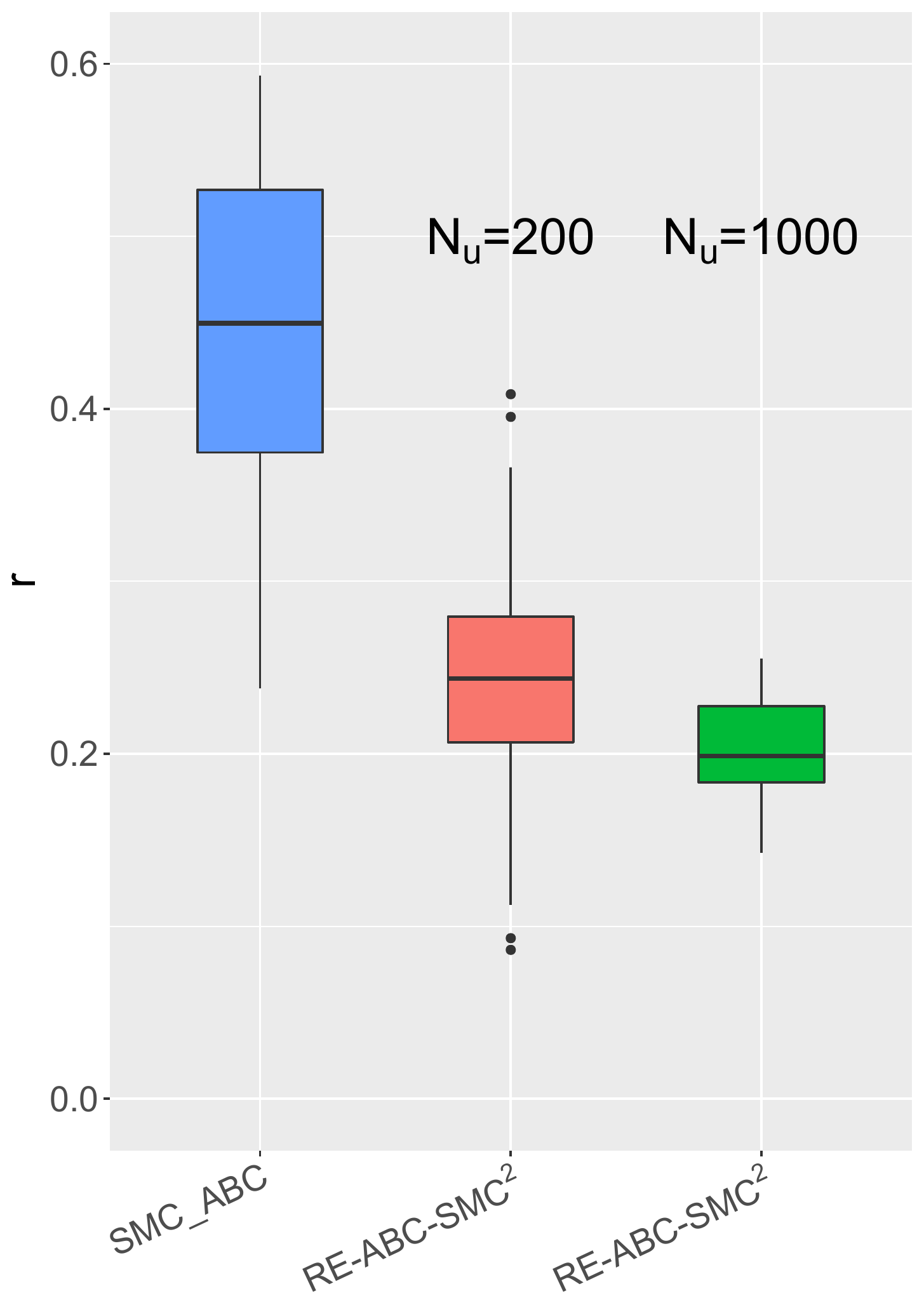} }}%
    \caption{Box plots of the estimated posterior means estimated by ABC-SMC and RE-ABC SMC$^2$ over 50 runs, for the two parameters: $p$ and right $r$.}%
    \label{fig:resmc2networkvariances}%
\end{figure}

We used $N_{\theta} = 500$ particles in our run of RE-ABC-SMC$^2$, with $N_u=200$. ABC-SMC was set up to have approximately the same cost per iteration as RE-ABC-SMC$^2$: we used $N_\theta=1.5 \times 10^6$ external particles, with $N_x=1$. Figure \ref{fig:resmc2networktol} gives a comparison of the adaptive schedules found using ABC-SMC and RE-ABC-SMC$^2$: again we see that for RE-ABC-SMC$^2$ the tolerance decreases more quickly, and a smaller tolerance in achieved. Figure \ref{fig:resmc2networkvariances} shows the box plots of posterior mean estimates from 50 runs of each algorithm, with parameter $p$ in the left figure and parameter $r$ in the right. The left box plot for each parameter corresponds to ABC-SMC, and the middle corresponds to RE-ABC-SMC$^2$ with the parameters described in the previous paragraph. Recall that the true value of $p$ is 0.5 and the true value of $r$ us 0.2. We observe that in all cases, the estimated posterior mean from RE-ABC-SMC$^2$ is closer to the true value with comparatively lower variance than that from ABC-SMC. We also again observe that the estimate from ABC-SMC is significantly biased, particularly for parameter $r$. Further, we ran RE-ABC-SMC$^2$ with an increased number of $N_u$ particles (the box plots on the right of each scenario) and once more observed an improved performance.

\section{Conclusions\label{sec:Conclusions}}

This paper builds on the particle MCMC methodology of  \citet{Prangle2016}, introducing an SMC counterpart to the approach in that paper. The advantage of the \citet{Prangle2016} approach over standard ABC techniques is that it uses an SMC, rather than an IS, likelihood estimator. IS requires an exponential number of points in the dimension of the data to control the variance of the likelihood estimator, rendering ABC impracticable unless the dimension of the data is reduced by considering only summary statistics. The rare event SMC approach requires only a quadratic \citep{Prangle2016} number of points in the dimension of the data, making it possible to use ABC on larger data sets without taking summary statistics. The new SMC$^2$ methodology in this paper inherits these advantageous properties, whilst also inheriting the useful properties of the widely-used ABC-SMC algorithms: that the particle population can be used to explore multi-modal targets \citep{Sisson2007f}; that it requires few tuning parameters, due to the adaptive techniques outlined in section \ref{sec:implementation}; and that the model evidence may be estimated directly from the SMC output \citep{didelot_likelihood-free_2011}. 

At the core of the approach is the use of the reparameterisation trick described in section \ref{sec:unpacking}. The real significance of this idea is that it removes the intractable likelihood from ABC, enabling one to use any available Bayesian computation technique for inferring the joint posterior of $\theta$ and $u$. This paper, and \citet{Prangle2016}, have in common that they use a pseudo-marginal approach for this task. However, just as outside the ABC context, the most appropriate technique depends on the structure of the posterior: in some cases it may be more effective to use the HMC of \citet{Graham2017}; in others it may be more effective to use an SMC directly on the joint posterior of $\theta$ and $u$ (as in a parallel work to ours in \citet{zhang_improvements_2022}). The use of a pseudo-marginal approach, and hence the algorithm in this paper, is effective when there is a strong posterior dependence between $\theta$ and $u$ and it is not easy to construct an MCMC on this joint space, and when one has available an effective approach to estimating the marginal (with respect to $u$) likelihood of $\theta$ (as is the SMC method used in this paper).

The limits of this reparameterisation trick are that in practice it may be difficult to rewrite a simulator model in this way. Often ABC is used in the context when ``unpacking the black box'' simulator is problematic for practical reasons: e.g. a model is developed over several years by a researcher in an applied field and is coded in such a way that it is not practicable to recode it in such a way that the sources of randomness $u$ can be isolated as inputs to the simulator. In this case, from an idealistic perspective it would be possible to use the methods from this paper, but in practice this would require substantial effort that may not be available. Another situation is the case mentioned in this paper, where the space of the some of the variables $u_r$ is sufficiently complicated that cannot (currently) hope to design an MCMC move that can explore the space effectively. For these variables the situation remains the same as for the original formulation of ABC: that the only available proposal is the simulator.

\section*{Acknowledgements}

Richard Everitt's work was supported by NERC grant NE/T00973X/1, and Ivis Kerama's work was supported by EPSRC grant EP/L016613/1 (the Centre for Doctoral Training in the Mathematics of Planet Earth).

\bibliographystyle{mychicago}
\bibliography{smc_rare_event}

\end{document}